\documentclass{mn2e}
\usepackage{amsmath}
\usepackage{epsfig}
\usepackage{graphicx}
\newcommand{\kpc}{\,\mathrm{kpc}}

\def\beq{\begin{equation}}
\def\eeq{\end{equation}}
\def\bey{\begin{eqnarray}}
\def\eey{\end{eqnarray}}
\def\beqarray{\begin{eqnarray}}
\def\eeqarray{\end{eqnarray}}

\def\mpc{\,{\rm {Mpc}}}
\def\Mpc{\,{\rm {Mpc}}}
\def\kpc{\,{\rm {kpc}}}

\def\kms{\,{\rm {km\, s^{-1}}}}

\def\v200{V_{200}}

\def\L850{L_{850\rm \mu m}}

\input epsf

\title[]
{Probing the Slope of Cluster Mass Profile with Gravitational Einstein Rings:
Application to Abell~1689  }

\author[H. Tu et al.]
{H. Tu$^{1,2,3}$\thanks{Email:tuhong@shnu.edu.cn}, M.
Limousin$^{4}$, B. Fort$^{2}$, C.G. Shu$^{1,3}$, J.F. Sygnet$^{2}$,
E. Jullo$^{5,6}$ \and J.P. Kneib$^{6,7}$, J. Richard$^{7}$
\\
  $^1$Physics Department, Shanghai Normal University, 100 Guilin Road, Shanghai 200234,
  China\\
  $^2$Institut d'Astrophysique de Paris, 98bis Bvd Arago, 75014 Paris,
  France\\
  $^3$Shanghai Astronomical Observatory, 80 Nandan Road, Shanghai 200030,
  China\\
  $^4$Dark Cosmology Centre, Niels Bohr Institute, University of Copenhagen, Juliane Maries Vej 30, 2100 Copenhagen,
  Denmark\\
  $^5$European Southern Observatory, Alonso de Cordova, Santiago,
  Chile\\
  $^6$OAMP, Laboratoire d'Astrophysique de Marseille - UMR 6110 - Traverse du siphon, 13012 Marseille,
  France\\
  $^7$Department of Astronomy, California Institute of Technology, 105-24, Pasadena, CA91125,
  USA\\
  }

\date{Accepted ........
      Received .......;
      in original form ......}

\begin{document}

\maketitle

\begin{abstract}
The strong lensing modelling of gravitational ``rings'' formed
around massive galaxies is sensitive to the
amplitude of the external shear and convergence produced by nearby mass
condensations. In current wide field surveys, it is now possible to
find out a large number of rings, typically 10 gravitational rings
per square degree. We propose here, to systematically study
gravitational rings around galaxy clusters to probe the cluster mass
profile beyond the cluster strong lensing regions. For cluster of
galaxies with multiple arc systems, we show that rings found at
various distances from the cluster centre can improve the modelling
by constraining the slope of the cluster mass profile.  We outline
the principle of the method with simple numerical simulations and we
apply it to 3 rings discovered recently in Abell~1689. In
particular, the lens modelling of the 3 rings confirms that the cluster is
bimodal, and favours a slope of the mass profile steeper than
isothermal at a cluster radius $\sim 300 \kpc$. These results are
compared with previous lens modelling of Abell~1689 including weak
lensing analysis.  Because of the difficulty arising from the complex
mass distribution in Abell~1689, we argue that the ring method will be
better implemented on simpler and relaxed clusters.

\end{abstract}

\begin{keywords}
Gravitational lensing---Galaxies: Clusters: General---Dark Matter
\end{keywords}

\section{Introduction}

In recent years, the modelling of cluster mass distribution has been
progressively improved by (a) coupling the strong lensing (SL)
analysis in cluster cores with weak lensing (WL) measurements at
large radius (e.g Gavazzi 2005\nocite{gavazzi05}; Limousin et al.
2007b\nocite{limousin1689}; Cacciato et al. 2006
\nocite{cacciato06}); (b) SZ measurements (e.g. Zaroubi et al. 2000
\nocite{zaroubi01}; Dore et al.  2001\nocite{dore01}; Seren
2007\nocite{Seren07}), (c) joint modelling of the cluster X-ray gas
distribution (e.g. Mahdavi et al.  2007\nocite{Mahdavi07}), (d)
dynamical analysis of the velocity distribution of the stars near
the potential centre (e.g. Miralda-Escud\'{e}
1995\nocite{miraldababul95}; Sand et al. 2004 \nocite{Sand04};
Gavazzi 2005\nocite{gavazzi05} and Koopmans et al.
2006\nocite{Koopmans06}).  Despite these important improvements, it
is not fully proved that cluster Dark Matter (DM) distribution
closely follows the ``universal'' profile (Navarro, Frenk \& White
1996\nocite{nfw}, hereafter NFW) predicted by N-body numerical
simulations. For this profile, the DM space density is cuspy at the
centre ($r \propto r^{-1}$ for $r \la r_s$) and behaves as $r^{-3}$
outward. Measuring with great accuracy these two main
characteristics is challenging and
no consensus has arisen yet with the present day lensing
observations.  Recent analysis has shown that S\'{e}rsic
profile (Merritt et al.  2005\nocite{Merritt05}) is also fitting
the universal mass profile of numerical simulations. Importantly, the
mass profile has only a weak dependence on the
halo mass or cosmology, allowing to stack different measurements
together to improve their significance.

At large radius, analyses of the hot gas distribution from the X-ray
observations (Pointecouteau 2005\nocite{Pointecouteau05}; Schmidt
2007\nocite{Schmidt07}), as well as weak lensing (Kneib et al.
2003)\nocite{kneib03} seem to favour an NFW-like profile for galaxy
clusters.  For elliptical galaxies, Wilson et al.
(2001\nocite{Wilson01}) claimed that they were consistent with
isothermal profile out to about $1 \Mpc$. Also, for massive
elliptical galaxies with gravitational rings, Gavazzi et al. (2007)
\nocite{Gavazzi07} found that the WL slope of the mass density could
be $\propto r^{-2}$ out to $300 \kpc$.

In the core of mass concentrations, the actual
existence of a DM cusp predicted by simple numerical simulations is
more unclear.  For spiral galaxy halos, the existence of a
singular density profile is still a debate because rotation curves
are better explained by isothermal profiles with a core radius
(Salucci 2003\nocite{Salucci03}). Projection effects of non circular
star orbits in triaxial halos has been invoked, but only in a few
cases, to explain the linear increase of the velocity at the centres
of galaxies (Hayashi et al. 2006 \nocite{Hayashi06}). As a
consequence, various mechanical processes, such as gas cooling,
supernova and AGN feedback, binary super massive black holes,
dynamical friction of the gas outflow during AGN activities, were
investigated to explain the formation of a DM core radius (Peirani
et al. 2006\nocite{Peirani06} and references therein).  For galaxy
clusters, it is often argued that isothermal ellipsoid mass
distributions with a flat core could better match the gravitational
arcs geometry than NFW profiles (Sand et al. 2004\nocite{Sand04};
Gavazzi 2005\nocite{gavazzi05}; Gavazzi et al.
2003\nocite{Gavazzi03}).  X-ray observations generally do not help
much for this issue because, due to the limited spatial resolution
of X-ray telescopes, one can just place upper limits on the radius
of the smallest DM core, around $30-50\kpc$ (Chen et al.
2007\nocite{Chen07}).  In summary, the DM density profile is still
an open question, and high quality data are necessary to settle
these questions.

The deviation of light by masses is well described by gravitational
lensing effect deduced from general relativity theory. The exquisite
{\it Hubble Space Telescope} images (particularly from the now
defunct ACS camera) are providing the necessary lensing constraints
to model observed gravitational lensing systems and may directly
probe the existence of a ``universal'' density mass profile.
However, one has to struggle to fully take into account the many
observational parameters entering a lens modelling.

Firstly, it is difficult to assess the stellar mass contribution
because the stellar mass-to-light ratio ($M/L$) is generally badly
determined, as well as the number of sub-halos and their galaxy
occupation numbers (Wright et al.  2002\nocite{Wright02}). Secondly,
lens modelling only probes the projected mass distribution of lenses
along the line of sight which introduces degeneracies in the 3D
density profile if the mass profile is only determined on a small
range of radius. When testing parametric models of mass
distributions for a given DM condensation, these current
difficulties are only partly alleviated if we can probe the
projected mass at many different radius.  As an example, to
disentangle between a flat core or a cusp with strong lensing it is
not enough to detect and to analyse gravitational images very close
to the centre, the so-called demagnified central images (Gavazzi et
al. 2003\nocite{Gavazzi03}) or inner radial arcs (Mellier et al.
1993\nocite{Mellier93}; Comerford et al. 2006\nocite{comerford06}).
Even in such ideal cases, some information on the mass distribution
beyond the Einstein radius is also critically needed.  Beyond giant
arc radii, one can use the information in the distortion of singly
highly magnified arclets, in an intermediate shear regime ($2 < \mu
< 3$), also called flexion regime (Bacon et al.
2006\nocite{Bacon06}; Massey et al. 2007\nocite{Massey07}). However,
a generic difficulty similar to the one encountered in the
weak lensing needs to be overcome. We do not know the shape of
background sources and the flexion method must be used in a
statistical way. Only with the most deepest space-based
observations, it becomes possible to reach surface number density of
background galaxies large enough to conduct such analysis on a
single cluster (see recent work on Abell~1689 by Leonard
et~al.(2007)\nocite{Leonard07}, where they reach a density of
background sources equal to $\sim 200$ sources/arcmin$^2$).

From the above discussion, we understand the difficulty to conduct
an accurate measurement of the slope  of the 3D density profile at
large radial distances. Hence considering  new probes of cluster
mass profile is important.

In this paper, we propose a method to investigate the slope of
cluster mass profiles. Gravitational image systems, i.e. ``rings''
formed around galaxy cluster members, are used to analyse the slope
of the cluster's density profile. Nowadays, such rings can be
systematically searched with dedicated software (e.g. Gavazzi et al.
2007, in preparation; Cabanac et al. 2005). Here, we outline the method with three rings
detected around Abell~1689.  The main goal is to provide some
constraints on the cluster potential at the location of the rings
(i.e. at $\sim$ 100$\arcsec$ from the centre of the brightest
cluster galaxy).  The coordinate system in this work is centred on
the brightest cluster galaxy: $\alpha_{J2000}=13:11:29.52,
\delta_{J200}=-01:20:27.59$.

The paper is organised as follows.  First we rapidly summarise the
properties of gravitational rings observed in the field around
elliptical lenses.  In section~\ref{simulation}, we illustrate the
method by using simulated cluster profiles, lensing galaxies, and
resulting images. These simulations show that we can put constraints
on the local slope of the projected mass distribution. In
section~\ref{reallife}, the method is applied to Abell~1689. In this
case we show that the three rings confirm that the cluster is
dominated by a bimodal mass distribution and that the local slopes
of both clumps are not much steeper than isothermal. Then, the
results are discussed relatively to previous lensing models of
Abell~1689, including a weak lensing analysis in the field of the
rings. Finally, we conclude that this method should be better used on very
relaxed clusters (single halo) with regular geometry to better probe
the slope of the mass profile at various distances from the cluster
centre.

Throughout this paper we assume a cosmological model with
$\Omega_{\rm{m}}$=0.3, $\Omega_\Lambda$=0.7, $H_0=70$km s$^{-1}
\Mpc^{-1}$. At the redshift of the cluster Abell~1689 ($z=0.185$) $1
\arcsec$ is equivalent to 3.089\kpc.

\section{Rings around elliptical galaxies}

The number of gravitational rings in the sky is large
(Miralda-Escud\'{e} \& Lahar 1992\nocite{Escude92}; Blandford
2000).Observations of strong lensing in the COSMOS field (Faure et
al. 2007\nocite{faure07}) have confirmed the estimation of their
surface number density in optical survey (about ten per square
degree with an average Einstein radius of about $0.8-1.5\arcsec$).
With the development of large surveys like the SDSS (Sloan Digital
Sky Survey) and more recently the CFHTLS
(http://www.cfht.hawaii.edu/Science/CFHLS/), several hundreds of
such arc systems are currently being discovered (Bolton et al.
2006\nocite{Bolton06}; Treu et al. 2005\nocite{Treu05}; Cabanac et
al. 2005\nocite{Cabanac05}). Almost all of them are multiple images
of background galaxies found around massive ellipticals. Robot
softwares are finding many of them among millions of objects in wide
field optical surveys and can then be studied in details with
dedicated follow-up with HST or adaptive optic systems (Marshall et
al. 2007\nocite{marshall07}). Future space-borne surveys like JDEM/SNAP
will allow us to discover as many as tens of thousands (Marshall
2005\nocite{Marshall05}).

Several structural properties have already been derived from the
studies of elliptical lenses detected in the SDSS survey (Koopmans
et al.  2006\nocite{slacs3}; Treu et al. 2006\nocite{slacs2}). It
has been found that 1) the stellar mass is dominant within the
Einstein radius and the velocity dispersion of stars almost matches
the velocity dispersion of the lens model ($\sigma/\sigma_{lens}=
1.01\pm0.02$ with 0.065 rms scatter); 2) the orientation of DM
coincide with the light distribution within the Einstein radius; and
3) the best fit for strong lens models coupled to a dynamical
analysis of the star velocity dispersion demonstrates that an
isothermal profile with $\rho\propto r^{-2 \pm 0.13}$ can describe
the total mass distribution at the Einstein radius.

For elliptical galaxies in clusters, one can expect that the light
distribution does not provide such a good geometrical description of the
mass distribution. According to simulation (Limousin et al.,
2007c\nocite{limousin07c}), the stellar contribution to the total
mass is larger for galaxies in the cluster core.  Indeed, as
galaxies are crossing the higher density region of the cluster, the
DM halo component is tidally stripped up to a radial distance which
is not much larger than the galaxy optical size. This halo stripping
has been studied observationally by galaxy--galaxy lensing
investigations. Although the deflection caused by galaxy scale mass
concentrations is small (i.e. shear $\gamma \sim 0.01$), it is
measured in many clusters (Natarajan et~al. 1998\nocite{Priya1};
Geiger \& Schneider 1999\nocite{geigeramas}; Natarajan et~al.
2002a,b\nocite{Priya2}\nocite{Priya3}; Limousin et~al.
2007a\nocite{limousin07a}) and there is a clear evidence for
truncation of galaxy dark matter halos in the higher density
environments. The inferred average half mass radius is found around
$40 \kpc$ in cluster cores, whereas half mass radii larger than $200
\kpc$ are derived for field galaxies of equivalent luminosity (see
Limousin et~al. 2007a\nocite{limousin07a}, for a review of
galaxy-galaxy lensing studies). Hence, the DM mass and stellar mass
can be of the same order within the ring.  However in the following, we will only consider
a total mass modelling of the ring.

The recent SL2S survey (http://www.cfht.hawaii.edu/
$\sim$cabanac/SL2S/) shows that the lens modelling of ``ring'' often
requires the contribution of an external shear that are likely
produced by nearby galaxy groups. In the outskirts of massive
cluster it is also very likely that ``rings'' are present and are
strongly affected by the cluster shear (Smail et al.
2007\nocite{smailcosmiceye}). For example in the HST/ACS images of
Abell~1689, several rings have been found (Limousin et al.
2007b\nocite{limousin1689}).

\section{Simulations of cluster rings}\label{simulation}

Cluster mass profiles can be described by different models such as
the NFW mass profile, a truncated cored isothermal model (Kneib et
al. 1996\nocite{kneib96}) or a cored power-law (Kneib et al.
2003\nocite{kneib03}; Broadhurst et al. 2005a). The latter two
models are defined by 3 parameters whereas NFW is only defined by 2
parameters.  For simple simulations we have chosen to describe the
cluster mass profile by a single dark matter clump parameterised by
a cored power law profile. The slope and core radius of the power
law are chosen in such a way that it can produce giant arcs at
$R_e\sim 45\arcsec$ similarly as in Abell~1689. Ideally, the radius
over which we will find rings should be significantly larger than
the core radius $r_c$ so that the slope of the mass profile is
almost constant at the radius of the ``ring''.  For a cored power
law, the 3D density can be written as:
\begin{equation}
\rho(r)=\rho_0(1+\frac{r}{r_c})^{-n}
\end{equation}
so that the logarithmic slope of the 3D mass density $\rho$ is given
by $n$.

Then, on top of this smooth cluster dark matter component we add a
galaxy at a distance $R$ from the cluster centre and we form a
``ring'' by lensing a distant galaxy. We then consider as
constraints \emph{only} the multiple images generated as part of the
ring(s). In order to reproduce the astrometric accuracy of the
HST-ACS images we add random errors for the position of the multiple
images ($\pm0.05\arcsec$ rms). In the following, these artificial
observations are then used to probe the slope $n$ of the cluster potential
without considering any additional giant arcs.
Different observational configurations are investigated with one
 similar to the Abell~1689 observation discussed later.

 We will investigate the results on the measurement of $n$ as a
 function of the ring distance to the cluster centre, and as a
 function of $n$.  All the simulations were done using the
 \textsc{lenstool} software (described in Jullo et~al.
 2007)\nocite{mcmc} that allows to easily investigate using a Bayesian
 approach the important model parameters and their degeneracies.

The rings in our
simulations are described by a truncated isothermal profile for
the total mass (PIEMD, Brainerd, Blandford \& Smail 1996\nocite{bbs})
\begin{equation}
\rho(r)=\frac{\rho_0}{(1+r^2/r_{core}^2)(1+r^2/r_{cut}^2)}
\end{equation}
with an ellipticity of the total mass
($ \epsilon = (a^2 - b^2)/(a^2 + b^2) $) and
position angle (PA) of the halo follow those of the light. In all
simulations the cluster centre is set at ($x=0, y=0$) coordinates.

\subsection{Case A: Single Ring Configuration}\label{seconering}

\subsubsection{Optimising ($n$)}

To investigate with which accuracy we can recover the slope of the
mass profile of the cluster, we have produced different mock
configurations by varying the slope ($n_{\mathrm{\textsc{in}}}$) and
generating multiple images around the lensing galaxy. Then we optimise
the slope of the cluster mass profile using these multiple images as
constraints to estimate ($n_{\mathrm{\textsc{out}}}$). With only one
``ring'', the maximum number of free parameters available is
three for a triple image system and five for a quadruple image system.

 For the ring we can either choose $\sigma$ or $r_{\rm{cut}}$. In
 principle, the velocity dispersion $\sigma_{\rm{ring}}$ can be
 measured from high resolution spectroscopy of the stellar component,
 so when there are only three parameters available, we thus choose to
 optimise $r_{\rm{cut}}$. Besides,
 the cluster slope $n$ and the velocity dispersion $\sigma_{cl}$ of the PL potential

\begin{equation}
\phi(r,\theta)=6\pi\frac{\sigma_{cl}^2}{c^2} \frac{D_{LS}}{D_S}r_0 (1+(r/r_0)^2[1+\epsilon \cos (2(\theta-\theta_0))])^{(3-n)/2}
\end{equation}
are the parameters we will optimise, as the other ones (centre,
 geometry, core radius) should be well known from the modelling of the
 cluster multiple images.
 As our preliminary goal is to see how well in this simplest
 simulation we can recover the input slope
 ($n_{\mathrm{\textsc{in}}}$) of the cluster we present the analysis
 for three different geometric configurations of the lensing galaxy:
 (a) galaxy halo without any ellipticity;  (b)galaxy halo with an
 ellipticity of 0.2 and PA=$0^o$ with respect to the
 line connecting the ring to the centre of the cluster (radial
 configuration) and (c)galaxy halo with an ellipticity of 0.2 and
 PA varying up to $90^o$ (orthoradial configuration).

\begin{figure}
\begin{center}
\includegraphics[height=8cm,width=8cm]{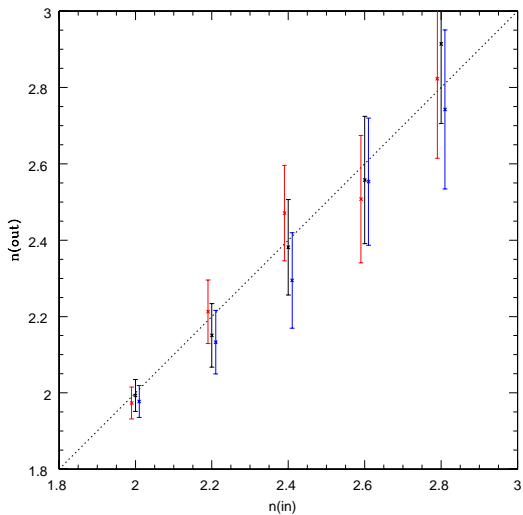}
\caption{ Results for a single ring configuration. The ring is
located 100$\arcsec$ from the cluster centre. The deformations are
generated using a power law profile whose slope is
$n_{\mathrm{\textsc{in}}}$ reported on the horizontal axis. Using
the constraints provided by the rings, we estimate this slope
$n_{\mathrm{\textsc{out}}}$ which is reported on the vertical axis.
The black points with error bars correspond to a sample of single
ring around a circular galaxy, red to a radial configuration, and
blue to an orthoradial configuration.  The black dashed line denotes
the equality. } \label{onering}
\end{center}
\end{figure}

Results of the simulations and modelling are shown in
Fig.~\ref{onering}. We have verified that the ellipticity and
orientation of the lensing galaxy relative to the cluster shear do
not change the accuracy in the recovering of the  slope $n$. We plot
the results for a fixed initial cluster velocity dispersion to show
that the shallower the profile, the better the recovery of the
slope. Indeed, the shallower the profile, the more massive the
cluster, and thus the stronger its influence on a ring galaxy
located at a fixed radius.

\subsubsection{Dependence with radial distance $R_{\mathrm{ring}}$}\label{secRring}

\begin{figure}
\begin{center}
\includegraphics[height=9cm,width=9cm]{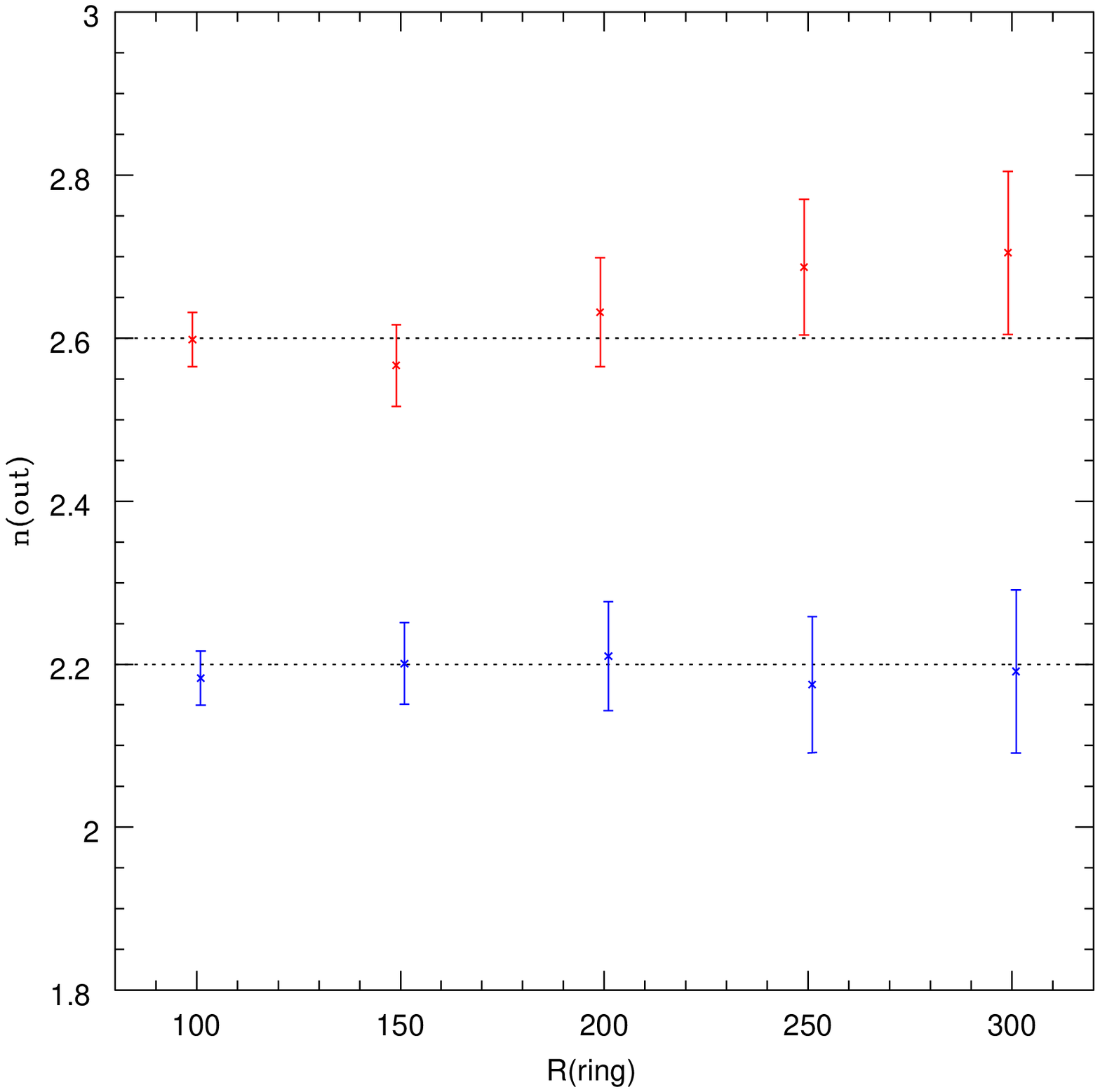}
\includegraphics[height=7.5cm,width=7.5cm]{f2b.ps}
\caption{(Top): dependence of the accuracy of the method with
respect to the distance between ring and cluster centre R(ring). In
red, the input slope is $n=2.6$, whereas in blue the input slope is
$n=2.2$. (Bottom): histogram of the probability density of $n$
values for the point 250 $\arcsec$ and $n=2.6$ taken from the top
figure.} \label{Rrings}
\end{center}
\end{figure}

When optimising $n$, we have fixed the rings to be at 100$\arcsec$
from the cluster centre because it corresponds to the radius of
rings we have found in the field of Abell~1689.   Finding rings
further away from the critical region, where the
shear of the cluster becomes weaker may bring interesting
constraints on the cluster mass profile. To investigate this
possibility with wider field survey of clusters, we have
conducted simple simulation varying the ring
radial distance from 100$\arcsec$ to 300$\arcsec$ to see how well we
can recover the local cluster mass profile. Results are shown in
Fig.~\ref{Rrings}.  The improvement in the determination of the slope
does not depend on the azimuthal distribution of the ring around the
cluster.  But indeed if several rings are aligned on a same direction, we can not
use them to estimate the cluster centre (see section~\ref{seccenter}).

At distance larger than $\sim$300$\arcsec$ ($\simeq 1\mpc$), as the
cluster influence is getting small, it is only possible to give a
lower limit on the slope $n$. Nonetheless, such a result can be
useful as it would show that the slope of the mass profile has
departed from the isothermal slope ($n=2$).  This can be understood
if we consider the total convergence $\kappa$ which is encompassed
within the ring radius R(ring). It comes from the
contribution of the galaxy and the cluster ($\kappa =
\kappa_{\rm{G}} + \kappa_{\rm{cluster}}$).  The convergence of the
galaxy $\kappa_{\rm{G}}$ cannot be arbitrarily high, also the
optimisation cannot accommodate any shallow cluster profile that will give a
cluster contribution $\kappa_{\rm{cluster}}$ which is too high,
since $\kappa_{\rm{G}} + \kappa_{\rm{cluster}}\simeq 1$.  The better
the likelihood on $\kappa_{\rm{G}}$ is, the better will be the
determination of $n$.  Also the star velocity dispersion within the
ring (total mass) can strongly improve the result.  Only at large
radius when both $\kappa_{\rm{cluster}}$ and $\gamma_{\rm{cluster}}$
become very small variation across the ring, the method
reaches its limit. (See Fig.~\ref{Rrings})

If a cluster has an NFW potential,  the logarithmic slope of the NFW
mass profile is
``continuously'' varying.  A power law approximation with a core can
match at the same time the total mass of the cluster within the ring
and the local slope, but it results in an unphysical core radius $r_{core}$.
Also for
real clusters and dataset, the best way is to directly test an NFW
model for the cluster component by including as many multiple images
as one can find (see section 4).

\subsubsection{Finding the cluster centre}\label{seccenter}
\begin{figure}
\begin{center}
\includegraphics[height=8cm,width=8cm]{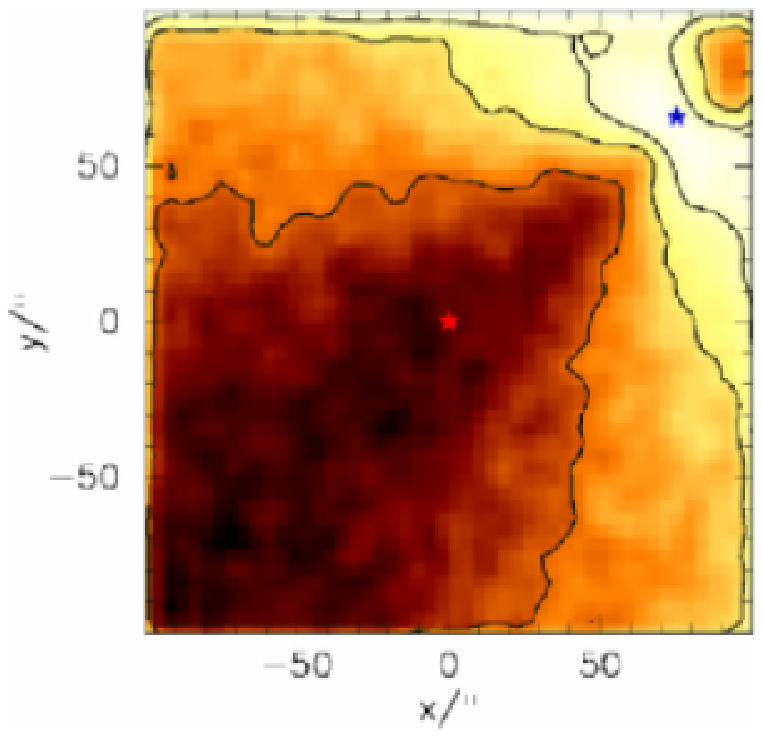}
\includegraphics[height=8cm,width=8cm]{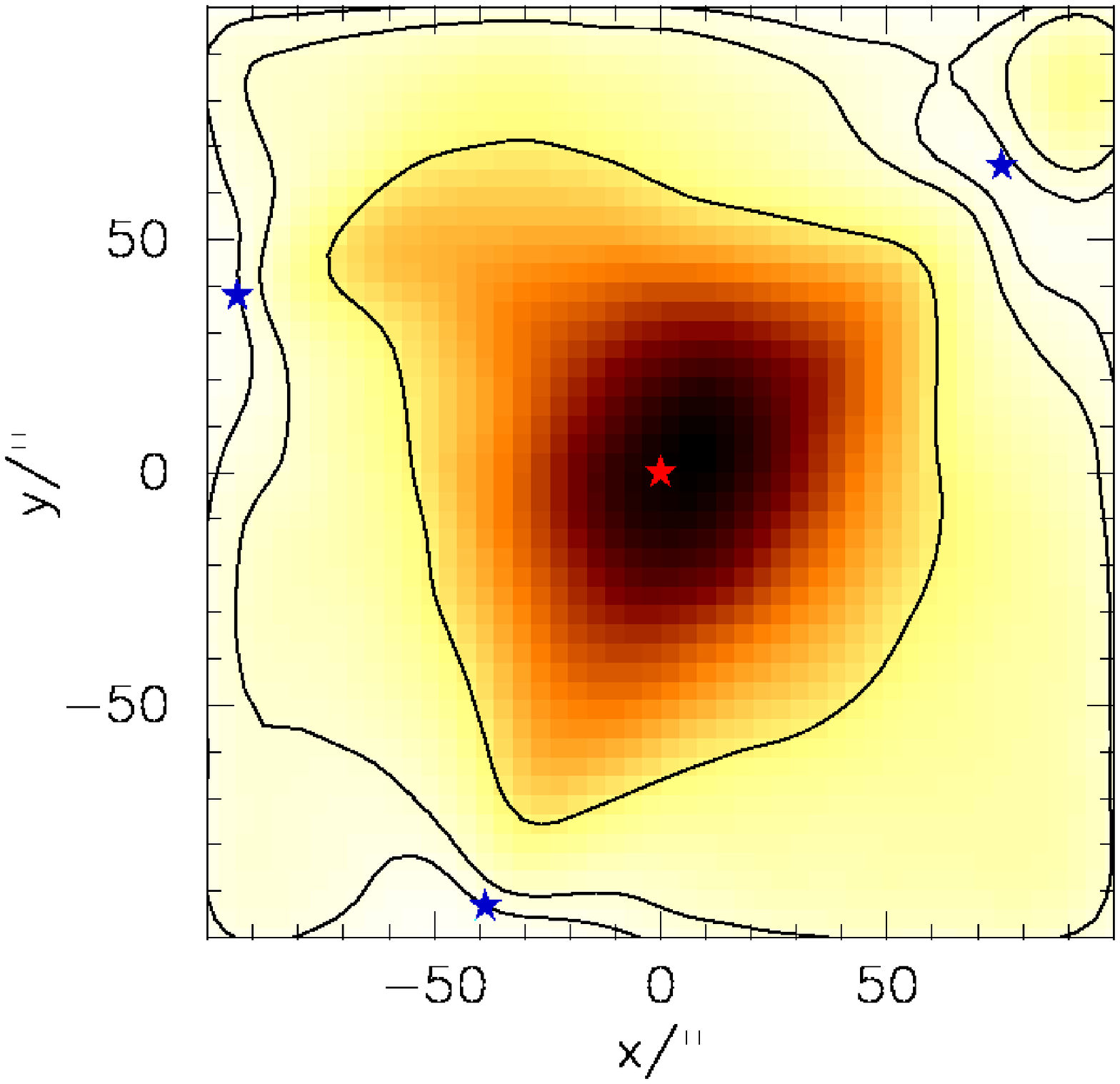}
\caption{Recovering the cluster position using the constraints
provided by the ring(s). (Top): one ring (labelled by a blue star) is
located at 100$\arcsec$ from the centre of the cluster ( red star).
(Bottom): similar results but with three rings (blue stars) located
at $\sim$100$\arcsec$.  The contours define the 1$\sigma$, 2$\sigma$
and 3$\sigma$ region.} \label{center}
\end{center}
\end{figure}

In this section, we investigate how to retrieve the location of the
cluster centre using several ring systems $\sim100\arcsec$ from the cluster
centre. In this exercise, we take the cluster position  as two free
parameters. As shown in Fig.~\ref{center}, a single ring cannot
reliably find the centre of the cluster but does give some
constraints on its direction. Note that the probability distribution
contours are large and not centred on (0,\,0).

\subsection{Case B: Multiple Rings Configuration}
We now consider three rings located at 100$\arcsec$ from a cluster
centre similarly as observed in Abell~1689. They surround the
cluster: one is in the north of the cluster, one is south and the
third one is west (see Fig.~\ref{center}).  We then conduct the same
simulations as done for a single ring but with the 3 rings.

\begin{figure}
\begin{center}
\includegraphics[height=8cm,width=8cm]{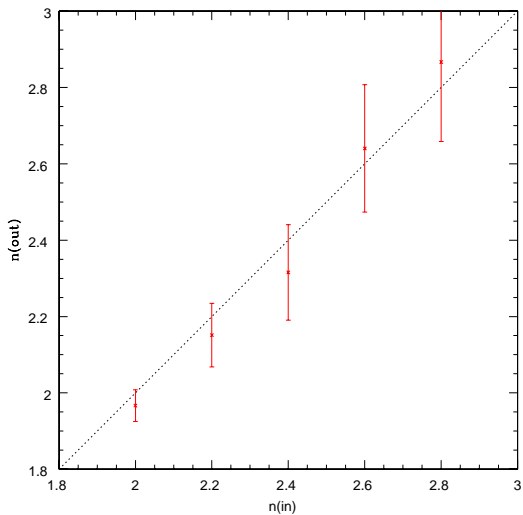}
\caption{Results for a three ring configuration similar to
Abell~1689.  The rings are located 100$\arcsec$ from the cluster
centre and surround the main clump.  The deformations are generated
using a power law profile whose slope is $n_{\mathrm{\textsc{in}}}$
reported on the horizontal axis. Using the constraints provided by
the rings, we measure the slope $n_{\mathrm{\textsc{out}}}$. }
\label{threerings}
\end{center}
\end{figure}

Results on recovering the cluster potential slope are shown on
Fig.~\ref{threerings}. Compared to Fig.~\ref{onering}, we see that
having three rings improves the measurement accuracy of the
slope following a $\sqrt{3}$ factor.

Regarding the cluster centre, we recover the centre of mass with the
following accuracy x=0$\pm$50$\arcsec$ and y=0$\pm$50$\arcsec$. The
constraints on the position of the clump are shown in
Fig.~\ref{center}.  This measurement is not very accurate comparing
to what we can achieve by analysing giant arcs systems generated by
a cluster. However, the simulation shows that not only are the rings
sensitive to the slope of the potential, but also if we have enough
rings, we can guess whether the location of the
 mass centre really dominated by a single massive clump.

\subsection{Conclusions from simulations}
A single ring can provide a good constraint on the slope of the cluster mass
profile beyond its Einstein radius because the modelling
constrains at the same time the local value of cluster convergence
and shear.  Schematically outlined, we can
say that the cluster convergence enlarges the Einstein radius of the ring
and the shear rotates and changes the ellipticity of the critical line.  Since
both the convergence and the shear depend on the slope $n$ at the location
of the rings ($R=R_{ring}$), the simultaneous adjustment of $\kappa_{cluster}$
and $\gamma_{cluster}$ provide a determination of $n$.  The ring method can
probe at which distance the slope of a cluster
potential departs from $n=2$ (isothermal slope).    Using simulations,
we are finding that there is a large range of cluster
radii ($R_e<R<10 R_e$) over which rings can efficiently probe the mass
profile slope.  An optimum case would be to observe rings at
various cluster-centric distances to better probe the overall DM
profile.

The ring method cannot put strong
constraints on the position of the clump perturbation.  The centre
of the clump should be determined independently to take full advantage of
the method (centre from X-ray or giant arc modelling).  However,
it can give information on the clumpiness
of the DM distribution since each ring is more sensitive to its closest
DM clump.  With many rings surrounding a single dominant cluster potential, one can measure more
accurately the slope of the mass profile, as well as determine the
centre of mass.

\section{Application to Abell~1689}\label{reallife}

With more than 31 arc systems, Abell~1689 (one of the richest
clusters of galaxies at intermediate redshift) is producing the
largest number of strong lensing images. Although the X-ray map
looks circular and is centred on the BCG, there is no cool core and
the gas has complicated dynamics (Andersson et~al.
2004\nocite{andersson04}). Furthermore, the line of sight velocity
dispersion is complex (Girardi et~al., 1997\nocite{girardi97}; Lokas
et~al. 2006\nocite{lokas06}). Broadhurst et al.
(2005b\nocite{tomsubaru}) found that the multiple arc systems
can be modelled with a power law potential with $n\simeq$3.08, a
surprisingly steep potential.  As we will show below the 3 rings do
not support a single power law potential.

In fact, the most recent strong lensing models show a bimodal mass
distribution (Miralda-Escud\'{e} \& Babul
1995\nocite{miraldababul95}; Halkola et~al.
2006\nocite{halkola1689}; Limousin et~al.
2007b\nocite{limousin1689}; Leonard et~al, 2007\nocite{Leonard07};
et~al. 2007). The cluster can be described as a dominant central dark matter
mass clump (O1 in the following) and a smaller perturbation
associated with the north-east galaxy group (O2 in the following).
Although this two mass clump description is not fully satisfactory
(see discussion in Limousin et~al., 2007b\nocite{limousin1689}), it
can be considered as a first order approximation for the
analysis presented here. Despite its complexity, we will
apply the method presented above on Abell~1689.
Furthermore, Abell~1689 has been extensively studied,
allowing us to compare our results to former studies as
a check for the ring method.

In the following analysis we will explore if the constraints
provided by these three rings can probe the bimodality of the mass
distribution, as well as the slope of each clump at the location of
the rings (about 100$\arcsec$ corresponding to $300\kpc$ from the centre)
without any strong a-priori assumptions coming from previous arc
modelling.  Next, we will use the results of previous modelling as input
parameters to check if they remain compatible with the ring configuration.

\begin{figure}
\begin{center}
\includegraphics[height=8cm,width=8cm]{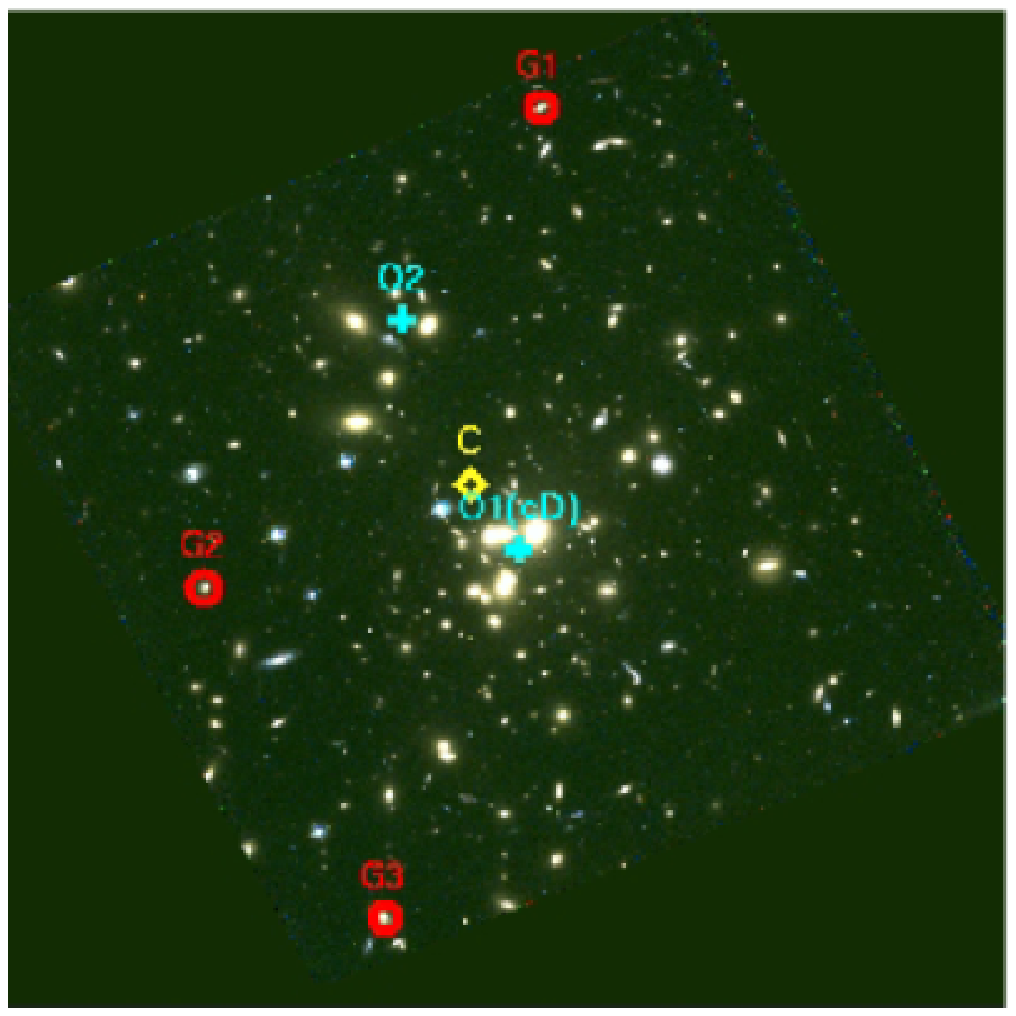}
\includegraphics[height=2.7cm,width=2.7cm]{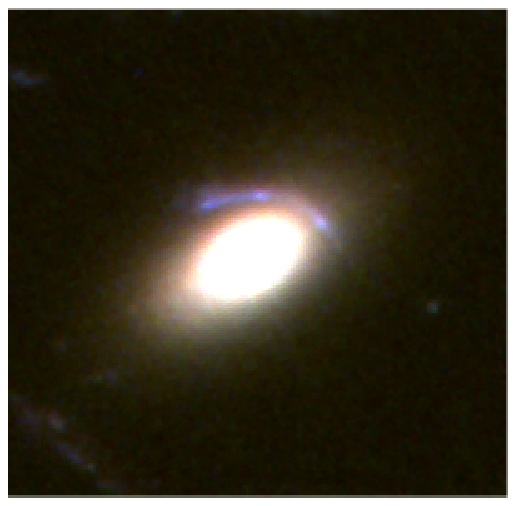}
\includegraphics[height=2.7cm,width=2.7cm]{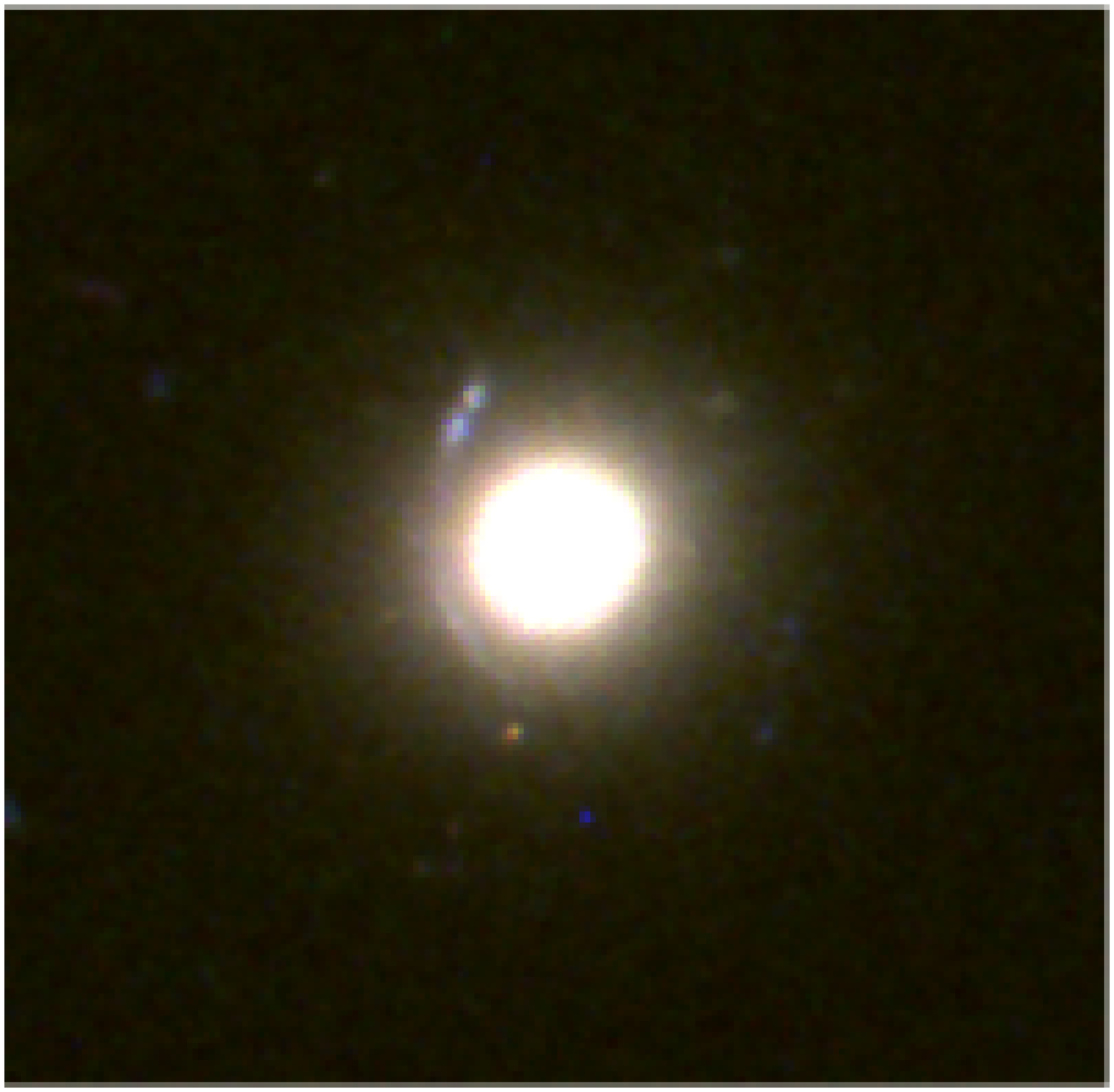}
\includegraphics[height=2.7cm,width=2.7cm]{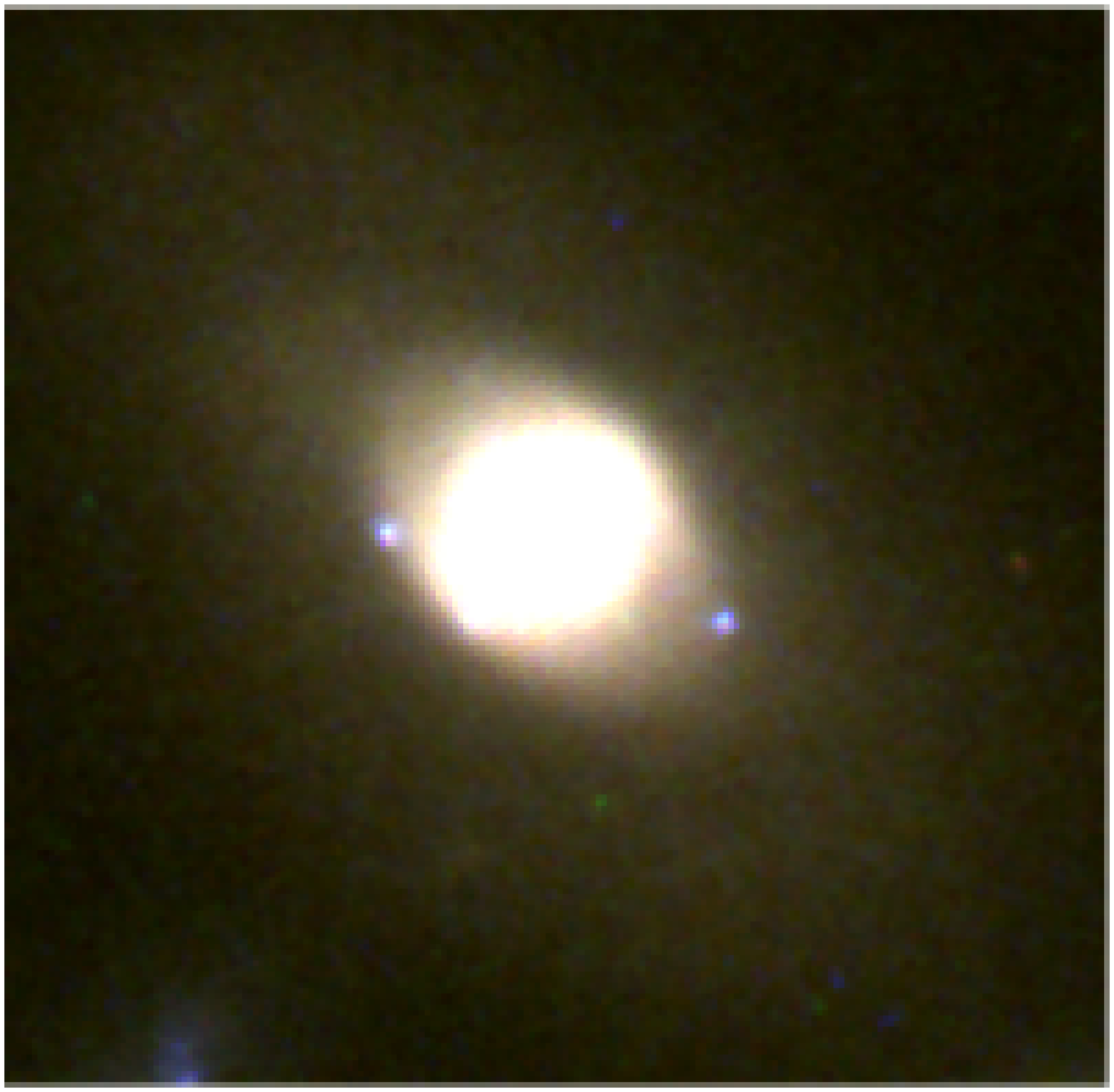}
\includegraphics[height=2.7cm,width=2.7cm]{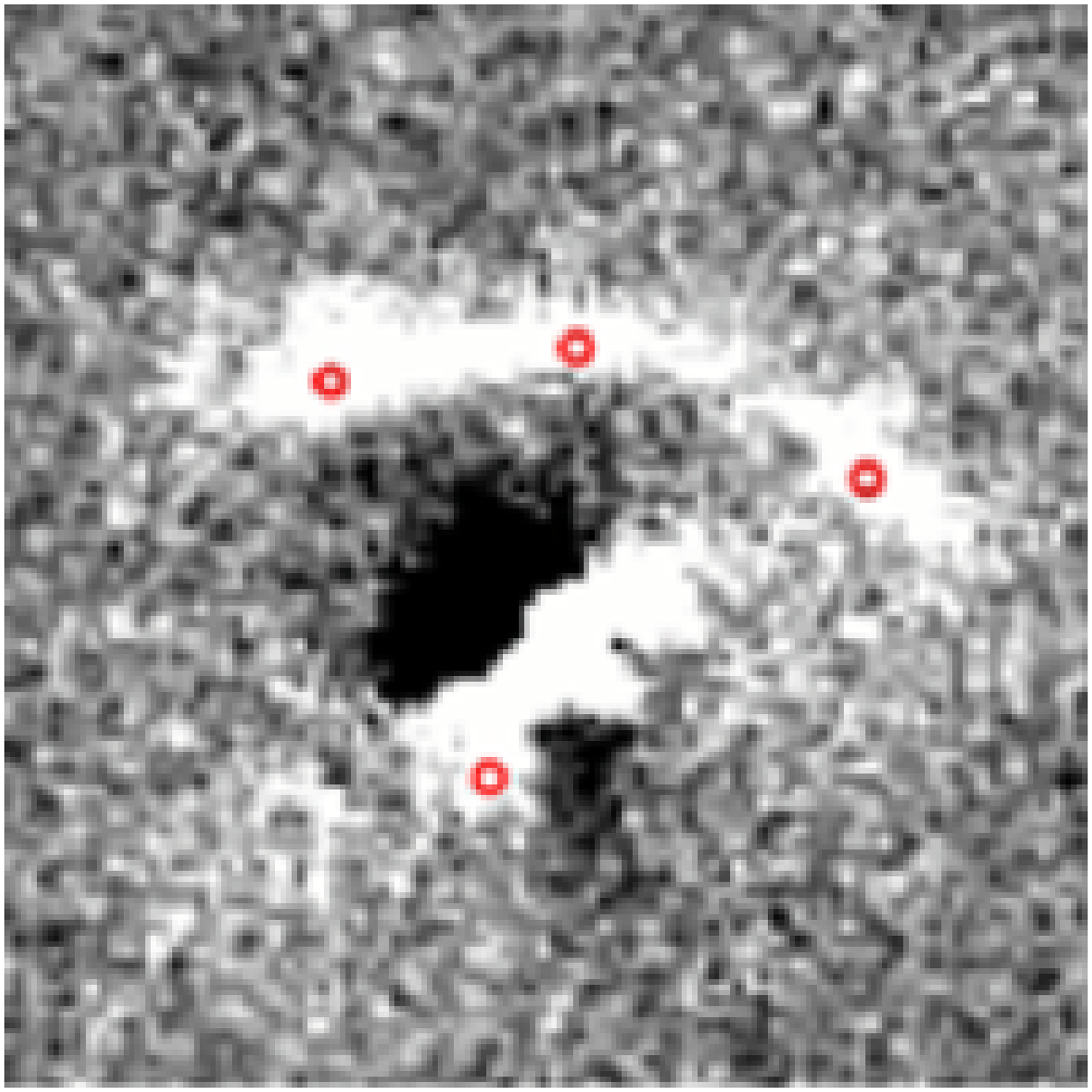}
\includegraphics[height=2.7cm,width=2.7cm]{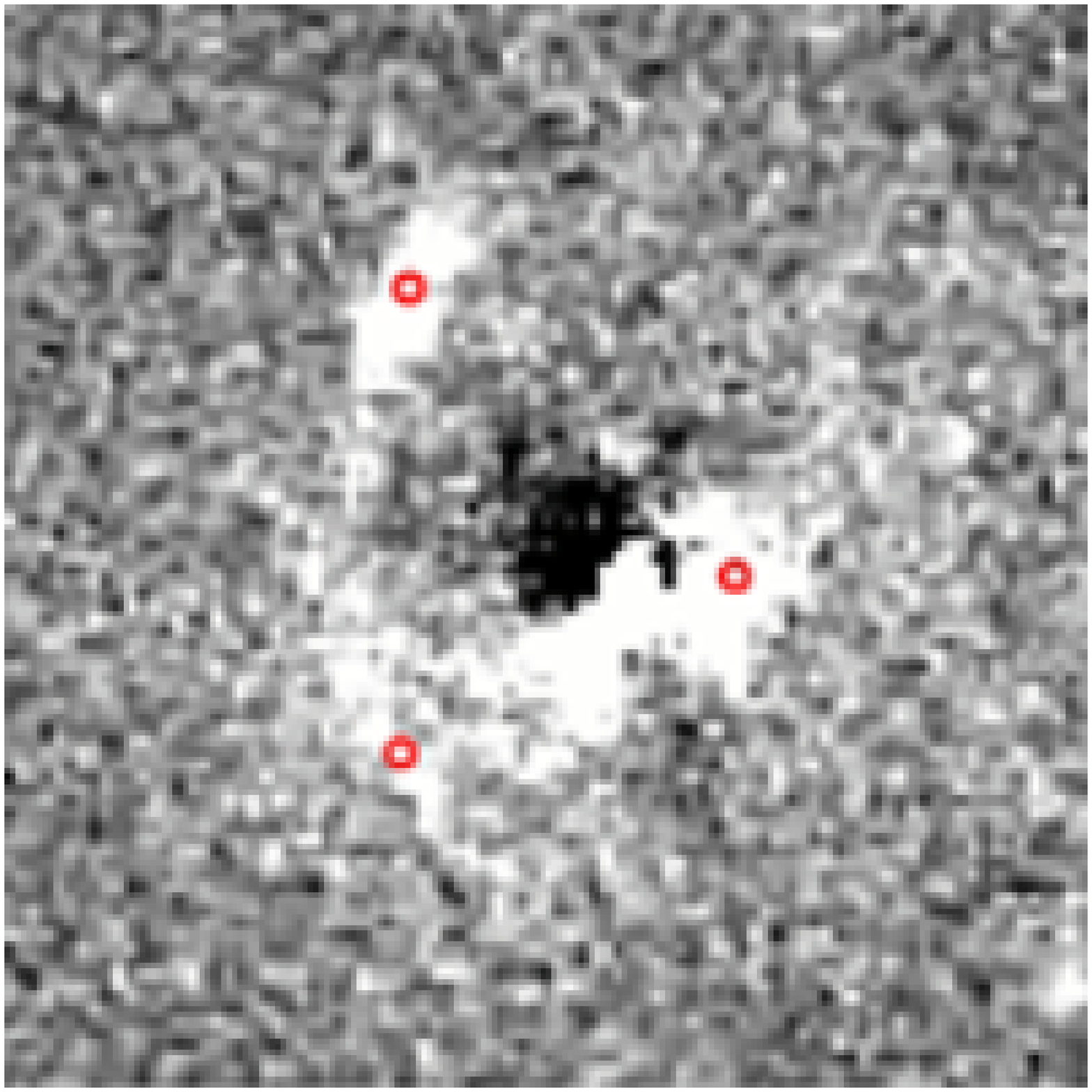}
\includegraphics[height=2.7cm,width=2.7cm]{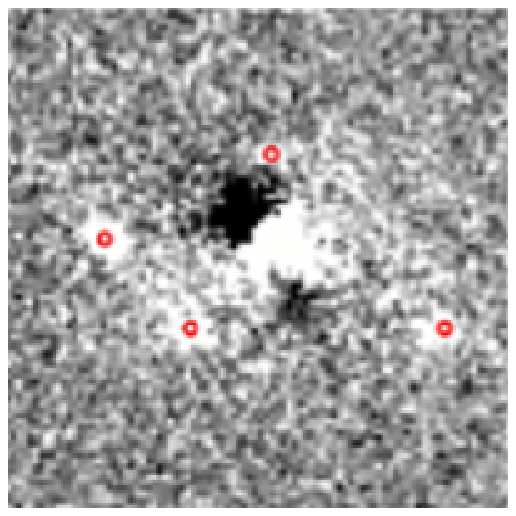}
\caption{Galaxy cluster Abell~1689.  The field size is $280\arcsec
\times 280\arcsec$, corresponding to $1089\kpc \times 1089\kpc$. G1,
G2 and G3 are the three rings involved in this work. The images with
lens galaxies subtracted are in the bottom row. The images we use as
constraints are marked as little red circles. The diamond (C) is the
optimised centre of the potential in the model we find with a single
potential. It goes to the direction of the luminous clump in the
north east (labelled by a cross O2). The centre found by X-ray is
marked by a cross (O1).} \label{rings}
\end{center}
\end{figure}

\subsection{Ring constraints}

Figure ~\ref{rings} shows the whole ACS field of view with the
location of the three rings, as well as three panels corresponding
to each ring. The three rings in Abell~1689 are at
$\sim$100$\arcsec$ ($\simeq 300 \kpc$) from the cluster centre
surrounding the two main luminous components (Table ~\ref{gglens}).

The three lensing galaxies are cluster members (based on their
redshift) and have similar colour as the cluster red sequence. The
lensed blue features around the ring G3 have a measured
spectroscopic redshift of z=1.91 from LRIS at Keck (Richard et~al.,
2007). For the other lensed features around G1 and G2, we will
assume for simplicity a redshift z=1. As a consequence, the velocity
dispersion that we will derive has a loose meaning because of the
assumption on the lensed redshift. The constraints we are
considering here are only those derived from the blue lensed
features detected around each ring (see Fig.~\ref{rings}).  For a
system with $N$ multiple images, we get $2(N-1)$ constraints.
Therefore, these give us a total of 16 constraints
(Table~\ref{gglens}).

In the following we \emph{only} use these 16 constraints in order to
probe the cluster without any {\sl a priori} coming from the arc
systems. The \textsc{lenstool} reduced $\chi^2$ quantifies how well
we reproduce the location of the gravitational ring images for each
family of lens models.  To
describe each ring system, we use a PIEMD profile (Kassiola \&
Kovner, 1993\nocite{kassiola}; Limousin et~al.
2005\nocite{limousin05}) for which we fix the position, ellipticity
and position angle according to the light distribution. Then, we
optimise the velocity dispersion $\sigma$ and cut-off radius
$r_{cut}$ for each ring. For ring G3, we also optimise the core
radius $r_{core}$ to ensure a better match with this special lensing
configuration although this has only a very minor impact on the
results presented below. So in total, we use 7 free parameters to
describe the three rings. This let us enough degrees of freedom to
explore some cluster properties.

\begin{table*}
\begin{center}
\begin{tabular}[h!]{ccccccccccc}
\hline \hline \noalign{\smallskip} Number  & \textsc{ra} &
\textsc{dec} & $\mathrm{d_{\textsc{bcg}}}(\arcsec)$ & Lens $z_{\rm
spec}$ &
Lens magnitude & $R_e(\kpc)$ & Images $z$  & Number of constraints\\
\noalign{\smallskip}
\hline
\noalign{\smallskip}
1  & 197.872 & -1.309 & 114  & 0.1758 & 18.25$\pm$0.13 & 2.0$\pm$0.1 & 1.0(assumed) & 6 \\
\noalign{\smallskip}
2  & 197.897 & -1.345 &  90  & 0.1844 & 18.58$\pm$0.16 & 2.2$\pm$0.2 & 1.0(assumed) & 4 \\
\noalign{\smallskip}
3  & 197.884 & -1.369 & 110  & 0.1855 & 18.11$\pm$0.01 & 4.1$\pm$0.1 &1.91          & 6 \\
\noalign{\smallskip}
\hline \hline
\end{tabular}
\caption{Strong galaxy-galaxy lensing events involved in this work:
coordinates (J2000), distance to the \textsc{bcg}; spectroscopic
redshift measurements when available; F775W \textsc{ab} magnitudes
obtained from the surface brightness profile fitting by Halkola et
al. 2006; circularised physical half light radius in units of $\kpc$
(Halkola et al. 2006); redshift of images; number of constraints
provided by these images.} \label{gglens}
\end{center}
\end{table*}

\subsection{Bimodality of the Cluster Mass Distribution}

If the central mass distribution of Abell~1689 is bimodal, the ring
configuration will be different as compared to the one in a single
central potential.  The local external shear at each ring location
has another direction.  We have tried to simultaneously model the
three rings with only one single central potential of either SIS,
NFW or power-law profile for which we optimise the position of the
centre and the dynamical parameters. We find that the $\chi^2$ per
degree of freedom is always larger than 100!

During this inconclusive optimisation, \textsc{lenstool} favours the
centre of the potential to be between O1 and O2 (Fig.~\ref{rings}).
This is unlikely since we know from both the lens model of the arcs
and the X-ray map that most of the mass is centred on the BCG.

Thus, we introduced a second mass component in the model. We
fix all the parameters of the main mass component O1 to the values
found in previous modelling (centred on the BCG) and we let free the
location and the velocity dispersion of the second mass component.
Much better solutions are found in this case, with a reduced
$\chi^2$ smaller than 4. In previous parametric strong lensing
studies, the position of the second mass component is not well
constrained.  In this study we find a value (RA=13:11:31.360 $\pm$
7($\arcsec$), y=-01:20:04.66 $\pm$ 21($\arcsec$)) close to the light
centre of O2. This test confirms that the three rings are sensitive
to the bimodality of the cluster central mass distribution, thus
providing an extra evidence for bimodality in this cluster.

\subsection{The slope of the mass distribution}

\subsubsection{Cluster modelling and rings}
The three rings are sensitive to the Abell~1689 gravitational
field and in principle we can use them to constrain the local slope of the
potential at the location of the rings. With the constraints of $3$
rings, we cannot explore too many parameters.  Also we
first assume that each mass component O1 and O2 can be described by
a power-law profile with a core radius.   Thus, we fix the
position of clump O1 to coincide with its brightest galaxy, and
the position of clump O2 with the barycenter of its brightest galaxies. We also
fix the ellipticity and position angle according to the cluster arcs modelling
of Halkola et al. 2006 (table~\ref{t2}).  The only free parameters
which are left free are the slope and velocity dispersion for each mass
clump (6 parameters for the two clumps). As discussed in section 4.1, we have 7
parameters to describe the three rings.  For the free parameters,
the ranges of variation
are chosen as follow.  The parameters of each
ring galaxy can vary between reasonable limits for a galaxy scale
potential, i.e. a central velocity dispersion $\sigma_0$ between
100$\kms$ and 300$\kms$, and the cut-off radius between 5 and 100
$\kpc$. The velocity dispersion of each
clump is allowed to vary up to $2500\kms$.
The slope $n$ can vary between 1.0 and 3.0. Under these hypotheses the slope
of the potential associated with the main central clump is found to be: $n=
2.4^{+0.1}_{-0.4}$, which is a little steeper than an isothermal sphere.
But surprisingly a large value of $108\pm32\arcsec$ is found for the core radius of O1.
It corresponds almost exactly to the distance of the closest rings.
Since we prefer to only use a power law approximation when the core radius
is really smaller than the radial ring distance
we try to force the core radius at the value
of Halkola et al. 2006 (table~\ref{t2}). Then we recover a similar value
$n=2.46^{+0.06}_{-0.60}$ for the main clump but the mass of the second
clump O2 becomes larger than the mass of O1.
It appears that a larger core radius can match the mass found for
each clumps with the modelling of the 31 gravitational arc systems
but a small core radius for O2 does not.
Clearly rings seem to tell us that the mass distribution of clump O2
cannot be approximated by a single central potential and that its mass distribution
is probably more complex. This can be understood if clump O2 is in fact
a projection of a filamentary distribution of mass along the line of sight.
We give as an illustration in Table~\ref{t2} the results of the
optimisation with the Halkola's
core radius values fixed (in brackets).
Besides, we will see below that
we have a degeneracy
of solutions between the velocity dispersion and $r_{cut}$ for each galaxy.

\subsubsection{Degeneracies}\label{sectdegeneracy}
Parametric strong lensing models with many parameters exhibit
various degeneracies. One reason is that strong lensing is sensitive to
the projected enclosed mass which is degenerated with respect to
some of the parameters describing the 3D mass density. For a discussion
on the degeneracies encountered in parametric lens modelling, one
can refer the work of Jullo et al. (2007)\nocite{mcmc}.

\begin{figure}
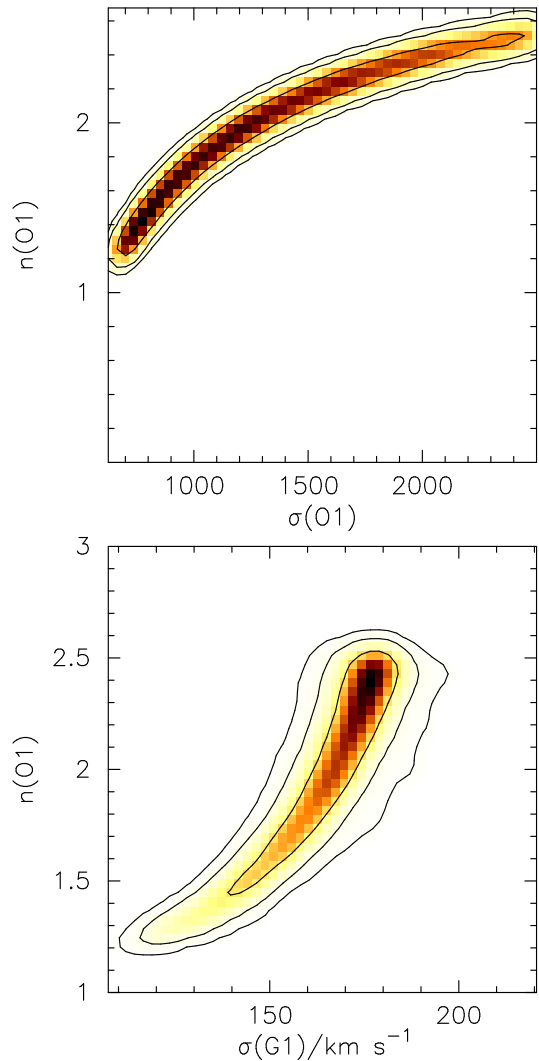

\begin{center}
\includegraphics[height=7cm,width=7cm]{f6a.ps}
\includegraphics[height=7cm,width=7cm]{f6b.ps}
\caption{Degeneracies between the slope of the main central
clump and dispersion of clump one and G1.  The peak value found for the slope
parameter is $n=2.46^{+0.06}_{-0.60}$ (see table~\ref{t2}).  These degeneracies
underline the importance in obtaining the velocity dispersion of
the clumps and lensed galaxies. }\label{degeneracies}
\end{center}
\end{figure}

In Fig.~\ref{degeneracies} we present the most relevant degeneracies
for the slope of the main clump O1 and the other parameters found in
this work.  We see that the velocity dispersion and slope $n$ of O1
are poorly constrained as expected.  Similar degeneracies are
obtained for O2. The bayesian framework in which \textsc{lenstool}
proceeds, it is important to specify the meaning of the ``best fit
model'', especially in such cases of strong departs from
Gaussianity. Throughout this paper we define the best fit value for
a given parameter as the mode of the marginalized distribution
(integrated over all other dimensions). Hence we can see some
unconsistancy between such values (summarised in Table~\ref{t2}) and
Fig. \ref{degeneracies} which directly comes from bayesian
techniques. For instance the high value for the slope
$n(O1)=2.4^{+0.06}_{-0.60}$ is essentially due to the choice of the
exploration range for $\sigma(O1)$.  The inferred value for $n(O1)$
would be lower when decreasing the upper limit of the variation
range of $\sigma(O1)$. For comparison, in a more frequentist
fashion, one would find values $n(O1)\simeq 1.5$ near the minimum
$\chi^2$. Fig.~\ref{degeneracies} also shows the interest of
measuring the velocity dispersion of G1 and clump O1 to break the
degeneracy and get valuable constraints on the density slopes.

We have noticed that the ring G2 is closest to O1.
Therefore, we make a test only with G2 to
see if it gives an equivalent result for the slope of O1.  Like the
single ring simulation in section~\ref{seconering}, only the
images around G2 are used in the optimisation.  Now the slope
and velocity dispersion of the main clump (O1) and the velocity dispersion
of G2 are optimised with only 4 constraints.
  In this case, we take other
parameters compatible with three ring modelling. A local slope is
found to be equivalent to $n=2.4\pm0.2$. For the second clump the
degeneracies are larger and we found the logarithmic slope could be
steeper than 2.4.

\begin{table}
\begin{center}
\begin{tabular}[h!]{c|ccc}
\hline \hline
\noalign{\smallskip} main clumps & Halo One  &  Halo Two & \\
\hline
\noalign{\smallskip} x($\arcsec$)    & [-2.1]       & [-33.1]      & \\
\noalign{\smallskip} y($\arcsec$)    & [-4.7]       & [56.7]       & \\
\noalign{\smallskip} $\epsilon$      & [0.159]      & [0.385]      & \\
\noalign{\smallskip} $\theta$        & [56.4]       & [106.3]      & \\
{\smallskip} $r_{core}(\arcsec)$     & [18.13]      & [24.28]      & \\
\noalign{\smallskip} $n$        & $2.46^{+0.06}_{-0.60}$  & $2.49^{+0.10}_{-0.27}$  & \\
\noalign{\smallskip} $\sigma$ (km/s) & $709.15\pm538.16$  & $1119.97\pm178.66$  & \\
\hline
\noalign{\smallskip} rings           & G1        & G2        & G3         \\
\hline
\noalign{\smallskip} x($\arcsec$)           & [3.527]   & [-86.088] & [-38.581]  \\
\noalign{\smallskip} y($\arcsec$)           & [113.703] & [-14.64]  & [-102.812] \\
\noalign{\smallskip} $\epsilon$      & [0.492]   & [0.09]    & [0.443]    \\
\noalign{\smallskip} $\theta$        & [-33.4]   & [-24.9]   & [43.8]     \\
\noalign{\smallskip} $r_{\rm{core}}$ ($\arcsec$) & [0.01]   & [0.01]   & $0.01 \pm0.07$    \\
\noalign{\smallskip} $r_{\rm{cut}}$  ($\arcsec$) & 14.6$\pm$16.9  & $<$ 10     & $<$ 10   \\
\noalign{\smallskip} $\sigma$ ($\kms$) & $173.74\pm13.99$  & $162.07\pm23.02$  & $174.70\pm27.40$  \\
\hline
\noalign{\smallskip} constraints     & & 16     & \\
\noalign{\smallskip} free parameters & & 11     & \\
\noalign{\smallskip} $\chi^2_{\rm{tot}}$   & & 18.27 & \\
\noalign{\smallskip} $\chi^2/d.o.f.$ & & 3.65  & \\
\hline
\hline
\end{tabular}
\caption{Best fitted parameters for the power-law mass distribution of the two clumps.
The parameters in brackets are fixed during the optimisation. }
\label{t2}
\end{center}
\end{table}

Ultimately, one would like to try a simultaneous modelling of the 3
rings together with the $31$ multiple arc systems of Abell~1689.
However, this would go beyond the scope  of this work for two
reasons. First, the description of the second clump is likely to be
more complex than previously thought.  Second, a comprehensive
investigation with \textsc{lenstool} would need an enormous amount
of computer time just to analyse one of several possible potential
configuration.  Instead, we can test published models of Abell~1689 to check
how well the rings are reproduced (we call this the ``ring test'').
To do this, we took as an input the structural parameters of the
cluster potential and we only solve for the ring configurations to
evaluate the $\chi^2$.  As we have already demonstrated above, the
power law model of Broadhurst et al. (2005a) does not pass the test
at all.  For the Halkola et al. (2006) models, the ring test gives a
reduced $\chi^2$ of 5.4 for the SIS model and 5.9 for the NFW model,
suggesting that the local slope for both case shall not depart much
from the isothermal solution.

\subsection{Conclusions for the Abell~1689 ring tests}
A preliminary conclusion is that modelling of the 3 rings seems to
favour slope a bit steeper than the SIS value at a radial distance
of 105$\arcsec$, with $n=2.4^{+0.06}_{-0.60}$. On the contrary, the
best models of the multiple arc systems (current Einstein radius of
about $50\arcsec$) favour an SIS or an NFW model with a scale radius
$r_s$ below $40\arcsec$.  Indeed, if $r_s$ is not significantly
larger than the radius where most of the tangential arc(let)s are
formed, the NFW slope remains close to $n=2$ when $r=r_s$ thus very
similar to an SIS model. Abell~1689 rings are at only two times the
cluster Einstein radius and are thus in a situation where we do not
expect a strong departure from an SIS slope. Hence the ring radius
R(ring) is not large enough to derive stringent conclusion.
Furthermore, the potential complexity of Abell~1689 is likely
increasing the degeneracies of the solutions.  In the following we
will try to discuss these results and their compatibilities with a
weak lensing analysis of Abell~1689.

\subsection{Discussions}
The lensed features around the three rings contain extra information
on the mass distribution and mass profile of the cluster at a
distance larger than the tangential arcs.  Rings confirm the Abell~1689
bimodality, they also show that the slope of the mass
distribution depart from isothermality at 100$\arcsec$ from the
cluster centre.

\begin{figure}
\begin{center}
\includegraphics[height=8cm,width=8cm]{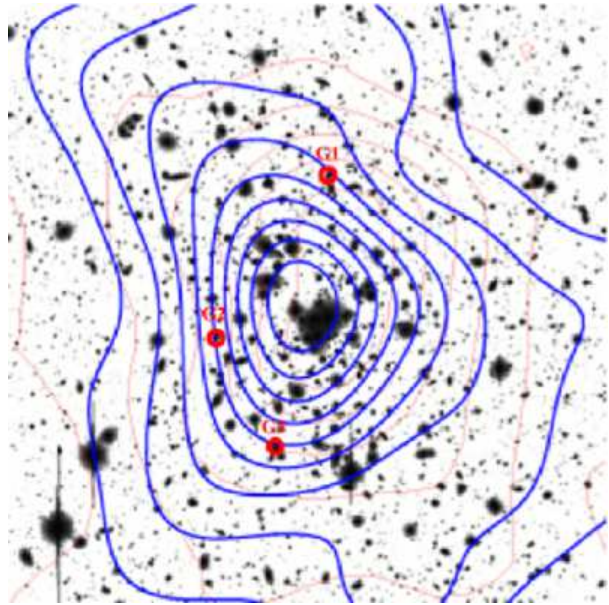}
\caption{The $\kappa$ map of Umetsu.  The field size is 15$\arcmin$
on a side.  Blue contours are the reconstructed lensing convergence
$\kappa$.  The lowest contour and the contour intervals are 0.05.
The 3 rings used in this paper are labelled with red circles.  Note
that the local slope of $\kappa$ is remarkably compatible with the slope
derived from the ring analysis.}
\label{umetsu}
\end{center}
\end{figure}

The three rings are located at $R_{\rm{ring}}\sim 105\arcsec$ from
the cluster centre. Considering that the dominant cluster mass
profile can be characterised by the parameters of O1 only, which is
legitimate given the difference in velocity dispersion between each
component, we find the slope of the 3D mass density to be equal to
$-2.4^{+0.06}_{-0.60}$. Describing Abell~1689 with an NFW profile,
this sets an upper limit on the scale radius $r_s$ of the NFW
profile: $r_s < R_{\rm{ring}}$. This upper limit on the scale radius
can be compared with the weak lensing analysis of Abell~1689.
Limousin et~al., 2007b found both the strong and the weak lensing
regime to agree in this cluster. From the CFHT data, they pursued a
weak lensing analysis, and found a scale radius $r_s = 93\arcsec$.
This value agrees with the upper limit provided by the ring
constraints. We can eventually try to go a bit further in this
comparison between the ring results and the shear map of Abell~1689.
Umetsu et al. (2007\nocite{umetsu}\nocite{okura07}) have obtained
deep images of Abell~1689 with the Subaru telescope and published a
projected mass reconstruction. From their figure 4 (See Fig
~\ref{umetsu}), we can see that their main centre is almost the
centre C determined by the 3 rings(Fig~\ref{rings}). Moreover, it is
possible to estimate from their map the local slope $n-1$ = $-\Delta
\mathrm{Log}(\kappa)/\Delta \mathrm{Log}(R)$ at the location of the
3 rings. We recover for G1, G2 and G3 an equivalent value of $n$
respectively equal to $1.95\pm0.2$, $2.48\pm0.1$ and $2.21\pm0.2$.
Although such a rough calculation of the slope is only valid for
power law mass distribution, it seems compatible with the ring
analysis. G2 is the closest ring to the centres of the two clumps
and the slope obtained from G2 ($2.4\pm0.2$, see
section~\ref{sectdegeneracy}) is similar to the Umetsu et al. (2007)
result. We can thus argue that a ring can bring similar information
as weak-lensing analysis but at a fixed point.

In summary, we found evidence for departure from isothermality at
100$\arcsec$ from the cluster centre (Fig~\ref{degeneracies}),
especially for the G2 location which is not perturbed by outside
sub-clumps.  Our results are not in disagreement with previous lens
modelling of Abell~1689. We found in particular that at the G2
location, the potential is steeper than isothermal.

The three ring galaxies do have a similar F775W magnitude, between
18.11 and 18.58. It is interesting to notice that the modelling
already infers comparable velocity dispersions values, around
170\,$\kms$, and small cut-off radii.  For rings G2 and G3, the
cut-radii are found smaller than $10 \arcsec$. These central
galaxies may be highly truncated, in good agreement with the tidal
stripping scenario (Limousin et al. 2007c\nocite{limousin07c},
Limousin, Sommer-Larsen et al. 2007\nocite{limousin07a}).  To
investigate further the exact value of cut-off radii, it is
necessary to wait for good measurement of the stellar velocity
dispersion within the ring radius.

\section{Conclusions}
This work revisits the importance of external shear perturbations
for the modelling of galaxy lenses (see Dye et al.
2007\nocite{Dye07}). But instead of focusing our attention on the
improvement of the ring modelling by introducing an external shear,
we probe the gravitational potential slope in the outskirts of
groups and clusters of galaxies thanks to the high sensitivity of
the ring modelling both to the local external shear and convergence of
the cluster.  We test the principle of
the method with simple simulations using power law models to
describe the cluster mass clumps.  From this simulation we found
that within the astrometric accuracy of ACS images a ring can be
used to detect a logarithmic slope of the external mass density
distribution up to about $n=2.8$. For an NFW profile with a typical
concentration parameter $C_{\rm{vir}} \sim 5$ (Comerford \&
Natarajan 2007\nocite{Comerford07}), this should allow to probe the
cluster potential up to a large fraction of its virial radius.  It
is now possible to detect rings in the outskirts of clusters with
new ground based surveys like CFHT-LS, but the determination of
the direction and amplitude of the external shear depends crucially
on the high resolution HST images. On the other hand, the knowledge
of the lens stellar velocity dispersion and of the redshift of the
ring itself would increase the robustness to the ring modelling (Koopmans
et al. 2006).

Using the 3 rings found in Abell~1689 we have shown that it is
possible to estimate the mass density slope.  Hence, we have shown
with the information given by rings alone that the potential of
Abell~1689 is bimodal. With this limited number of rings it is
however not possible to better trace other DM substructures within
each clump. Therefore a global modelling of the 31 arcs systems and
rings remains a good challenge.

We find a mass density slope a bit larger than $n=2$ for each DM clump at
about $100\arcsec$ from their centre, but the complexity of the
cluster potential really weakens the result. Since the most immediate
appealing application of the method is to probe the slope departure
from a $r^{-2}$ mass density profile, we strongly suggest similar analysis on
less complex clusters. The ideal lenses to probe the reality of a
universal NFW profile would be a cluster or a fossil group with a
bright cD galaxy having a single dominant halo, and which displays
at least a distant ring, a multiple arc with known redshift, and a
radial arc. Such a configuration would probe the potential at
various distances from the centre. The discovery of such ``golden
lenses'' requests a large survey of massive clusters. Most often
clusters with a great number of arcs have multi-polar potentials
(i.e. longer caustics) so that are more complex to analyse. However,
rings are influenced by all their nearby environment (including
foregrounds). In this sense, a statistical study of the
environmental effect on the rings should be conducted in the future.

Many rings can be discovered now in the field of deep wide field
surveys. It will be possible to improve the method and to use it
more systematically. As an example, the CFHT-LS ``wide'' survey will
cover 170 degree , and we expect to detect about 10 rings per square
degree with an average redshift $z_{\rm{lens}} = 0.65$ (Cabanac et
al. 2007\nocite{Cabanac07}). The clusters and groups in front of or
at the same redshift as the lensing galaxies have approximately the
same sky density (Oguri 2006\nocite{Oguri06}). Thus we can expect
that several rings per square degree will be influenced by the
external shear of a nearby mass condensation in the field. Such
cases are already observed in the SL2S survey (Cabanac et al.
2007\nocite{Cabanac07}).

For Abell~1689, we have found three rings within the small HST field
($r \sim 100\arcsec$), which may be consider as surprisingly high
number. In fact, the large number of ellipticals in this cluster and
the extra convergence that it adds to the lensing effect most
probably compensate for the small field of view imaged with the ACS
camera\footnote{After the submission of this paper, King (2007\nocite{king07})
 has shown with cluster simulation that the ring cross section is
increased by a factor about 3 near critical lines. }.  We also found two
possible rings closer to the critical region of Abell~1689 ({\it
i.e.} at radii smaller than 50$\arcsec$). However, we are not
considering these rings here, because these systems are in a crowded
region and can not be described by a simple contribution of a galaxy
and a cluster scale component. Any attempt of modelling these
latter rings must involve all the multiple arc systems of Abell~1689.
In wide field
surveys, perturbing clusters may be found at
a lower redshift than the ring-producing elliptical lenses (e.g.
Smail et al. 2007\nocite{smailcosmiceye}). In such cases, a proper
analysis requires a multi-plan ring modelling.

In conclusion, rings seem to be a promising tool to constrain the
mass distribution slope at large radii from the centres of groups
and clusters provided dedicated observations are carried out on a
few golden lenses. They can provide information on many structural
parameters of halos of clusters and galaxies. In complement to the
modelling of the multiple arcs in the cluster core, they can confirm
or otherwise dispute the existence of the universal DM halo profile
predicted by numerical simulation, as well as to study sub-halos and
their cut-off radii. For these studies, it is crucial to measure the
velocity dispersion of the lens galaxies which produce the
gravitational ring images. It clearly appears from the study of
Abell~1689 that this method should be implemented first on a sample
of very relaxed clusters or (fossil) groups in order to analyse a
single DM potential with the simplest possible geometry. Only when
large sample of rings is available, it will become possible to start
a systematic analysis of the external shear perturbation on ring
shapes in correlation with more complex nearby (eventually
foreground) environment.

\section*{Acknowledgement}
The authors are very grateful to Raphael Gavazzi for very helpful discussions
and comments on the paper. This
project is partly supported by the Chinese National Science
Foundation No. 10333020, 10528307, 10778725, 973 Program No.
 2007CB815402, and Shanghai Science Foundations. And this work was
 undertaken within the frame of the French ``Agence Nationale de la
 Recherche'' (ANR) project BLAN06-3$\_$135448. HT acknowledges the financial support of the
Shanghai Science Foundation for a visit to IAP and is grateful to
the hospitality during the visit. HT also acknowledges the Dark
 Cosmology centre for their invitation in Copenhagen where part of
 this work has been carried out.  The Dark Cosmology Centre is funded
 by the Danish National Research Foundation.   JPK acknowledges
support from CNRS.  This work is based on observations made with the
 NASA/ESA Hubble Space Telescope, obtained through the archives of
 the Space Telescope Science Institute, which is operated by the
 Association of Universities for Research in Astronomy, Inc., under
 NASA contract NAS 5-26555.

\bibliography{A1689}

\end{document}